\documentclass{jfm}

\usepackage{upmath}
\usepackage{amsmath}
\usepackage{amsfonts}
\usepackage{amssymb}
\usepackage{pgf,pgfarrows}
\usepackage{wasysym}
\usepackage{graphicx}
\usepackage[export]{adjustbox}
\usepackage{subfig}
\usepackage{natbib}
\usepackage{upgreek}
\usepackage{mathrsfs}
\usepackage{float}
\usepackage{caption}
\usepackage{soul}
\usepackage{textcomp}
\usepackage{stmaryrd}
\usepackage{color}
\usepackage{todonotes}
\ifCUPmtlplainloaded \else
  \checkfont{eurm10}
  \iffontfound
    \IfFileExists{upmath.sty}
      {\typeout{^^JFound AMS Euler Roman fonts on the system,
                   using the 'upmath' package.^^J}%
       \usepackage{upmath}}
      {\typeout{^^JFound AMS Euler Roman fonts on the system, but you
                   dont seem to have the}%
       \typeout{'upmath' package installed. JFM.cls can take advantage
                 of these fonts,^^Jif you use 'upmath' package.^^J}%
      }
  \else
  \fi
\fi

\ifCUPmtlplainloaded \else
  \checkfont{msam10}
  \iffontfound
    \IfFileExists{amssymb.sty}
      {\typeout{^^JFound AMS Symbol fonts on the system, using the
                'amssymb' package.^^J}%
       \usepackage{amssymb}%
         \let\leq=\leqslant
         \let\geq=\geqslant
      }{}
  \fi
\fi

\ifCUPmtlplainloaded \else
  \IfFileExists{amsbsy.sty}
    {\typeout{^^JFound the 'amsbsy' package on the system, using it.^^J}%
     \usepackage{amsbsy}}
    {\providecommand\boldsymbol[1]{\mbox{\boldmath $##1$}}}
\fi

\newsavebox{\astrutbox}
\sbox{\astrutbox}{\rule[-5pt]{0pt}{20pt}}

\def\ee{{\rm e}}
\def\ii{{\rm i}}

\affiliation{
$^1$ Environmental and Geophysical Fluids Group, Department of Mechanical Engineering, Indian Institute of Technology Kanpur, U.P. 208016, India.\\
[\affilskip]
$^2$ Departments of Aerospace and Mechanical Engineering, Civil Engineering, Mathematics, and Information and Operations Management, University of Southern California, 430K Olin Hall, Los Angeles, CA 90089-1453.
}
\pubyear{2012}
\volume{xxx}
\pagerange{xxx--yyy}
\title[Non-linear dynamics induced by linear wave interactions in multi-layered flows]{Non-linear dynamics induced by linear wave interactions in multi-layered flows}
\author[A.~Guha and F.~E.~Udwadia]
{
A\ls N\ls I\ls R\ls B\ls A\ls N\ns G\ls U\ls H\ls A$^{1}$\footnote{Electronic mail for correspondence: anirbanguha.ubc@gmail.com }
\and
F\ls I\ls R\ls D\ls A\ls U\ls S\ns E.\ns U\ls D\ls W\ls A\ls D\ls I\ls A\ls$^{2}$
}
\date{?? and in revised form ??}

\begin{document}

\maketitle

\begin{abstract}
 Using simple kinematics, we propose a general theory of linear wave interactions between the interfacial waves of a two dimensional (2D), inviscid, multi-layered fluid system. The  strength of our formalism is that one does not have to specify the physics of the waves in advance.  Wave interactions may lead to instabilities, which may or may  not be of the familiar ``normal-mode'' type. Contrary to intuition, the underlying dynamical system describing linear wave interactions is  found to be non-linear. Specifically, a saw-tooth jet profile with three interfaces possessing  kinematic and geometric symmetry is explored. Fixed points of the system for different ranges of a Froude number like control parameter $\gamma$  are derived, and their stability evaluated.   Depending upon the initial condition and $\gamma$, the dynamical system may reveal transient growth,   weakly positive Lyapunov exponents, as well as different non-linear phenomena such as formation of periodic and pseudo-periodic orbits. All these occur for those ranges of $\gamma$ where normal-mode theory predicts neutral stability.
Such rich non-linear phenomena is not observed in 2D dynamical system resulting from the $2$-wave problem, which only reveals stable and unstable nodes.
\end{abstract}

\section{Introduction}
Layered flows are often encountered in many geophysical and engineering problems. During  summer, sharp thermoclines in lakes and oceans typically divide warmer (lighter) water above from the colder (denser) water below \cite[]{woods1968}, thereby producing an approximately ``two layered'' system. 
Zonal jets, consisting of layers of nearly constant potential vorticity, are ubiquitous in the terrestrial atmosphere and in the oceans, as well as in the atmospheres of the gas giant planets \cite[]{scott2012}. Multi-layered Poiseuille flows are often encountered in engineering, especially during co-extrusion, lamination and coating processes \cite[]{moyers2004}. An interface separating two neighbouring layers supports  neutral progressive wave(s). For example, the interface between air and water supports surface gravity waves, while that between cold and warm water supports interfacial gravity waves. A fluid flow can become unstable when multiple interfaces are present. The ensuing instability can potentially cause transition to
 turbulence, a problem of immense importance in nearly all sub-fields of fluid dynamics. 
 
 Normal-mode instabilities in homogeneous and density stratified shear layers (e.g.\ Rayleigh/Kelvin-Helmholtz, Holmboe, Taylor-Caulfield instabilities) can be explained through resonant interaction between two interfacial waves \cite[]{tayl1931,bret1966,caul1994,bain1994,heif2005,guha2014}. Recently \cite{guha2014} (hereafter GL14) proposed a generalized theory of two interacting linear
waves,  known as the ``Wave Interaction Theory (WIT)''. WIT  adds to the mechanistic understanding of  normal-mode shear instabilities. According to WIT shear instabilities arise due to \emph{synchronization} of two interfacial waves (and not simply due to resonance).  Drawing analogies from coupled oscillator synchronization, WIT extends the wave interaction formalism to accommodate non-normal (or non-modal)  instabilities as well. It reveals that, due to non-normality, shear instabilities can lead to large transient growths in interfacial wave amplitudes, often  surpassing normal-mode growth by few orders of magnitude. Standard linear stability theory based on  normal-mode ansatz would fail to capture this behaviour. GL14  showed that such large growth could arise if the normal-mode ansatz is not imposed on the governing PDEs. They found that the underlying dynamical system describing the interacting wave amplitudes and phases is highly non-linear, which explains the reason behind large transient growths. Although transient growth mechanism due to non-normality is well understood \cite[]{tref1993,schmid2001stability}, WIT provides a simple mechanistic explanation  in a minimal setting with two waves.

 The main goal of this paper is to study linear instabilities that arise via multiple wave interactions \emph{without} limiting the analysis to the   normal-mode formalism. Unfettered by the conventional normal-mode ansatz, both normal-mode and non-modal instabilities are thus explored. WIT theory has so far been limited to the  interaction between just two linear interfacial waves. While two wave interaction provides the mechanistic picture of well known shear instabilities, there would arise many physical scenarios in the oceanic and atmospheric systems where the use of just two interfaces (or waves) could be an unrealistic over-simplification. Moreover, the phase-portrait of two wave WIT is analogous to coupled oscillators and is therefore very simple. For wavenumbers satisfying ``linearly unstable'' criteria, the phase-portrait exhibits two fixed points, one is a stable node (growing normal-mode) while the other is an unstable node (decaying normal-mode) \cite[]{heif1999,guha2014}. One can therefore anticipate richer non-linear dynamics when  multiple interfaces are considered. 
 This paper deals with developing a  multi-interface framework and investigating the resulting dynamical system.  As shown, the extension from two interfaces to multiple interfaces turns out to be quite non-trivial. Such multi-layered systems are themselves often  idealized models of real-world fluid systems. In reality quantities of interest vary continuously;  modeling base states, which are continuous functions, as piece-wise continuous (which is needed for multi-layered systems) is indeed a simplification. Yet, such simplifications often help, and in many instances are indeed necessary for  providing the needed analytical tractability in order to develop improved insights and useful results. For example, in (homogeneous) shear flows, the base-flow vorticity varies continuously, but for greater analytical tractability it can often be assumed to be layer-wise constant. Likewise, flows in the atmosphere and the oceans are often modeled as  multi-layered shallow-water systems since this provides a simplified representation, while retaining their key dynamical features \cite[]{vallis2006}. In this paper we first develop  a general framework for multi-layered systems. Then we specifically consider and provide computational results for the ``saw-tooth'' jet problem, which   is an approximate model for the zonal-jet structure in planetary atmospheres. Furthermore, the saw-tooth jet  exhibits  three interfaces and possesses kinematic and geometric symmetry.
 
  \begin{figure}
\subfloat[]{\includegraphics[width=2.6in]{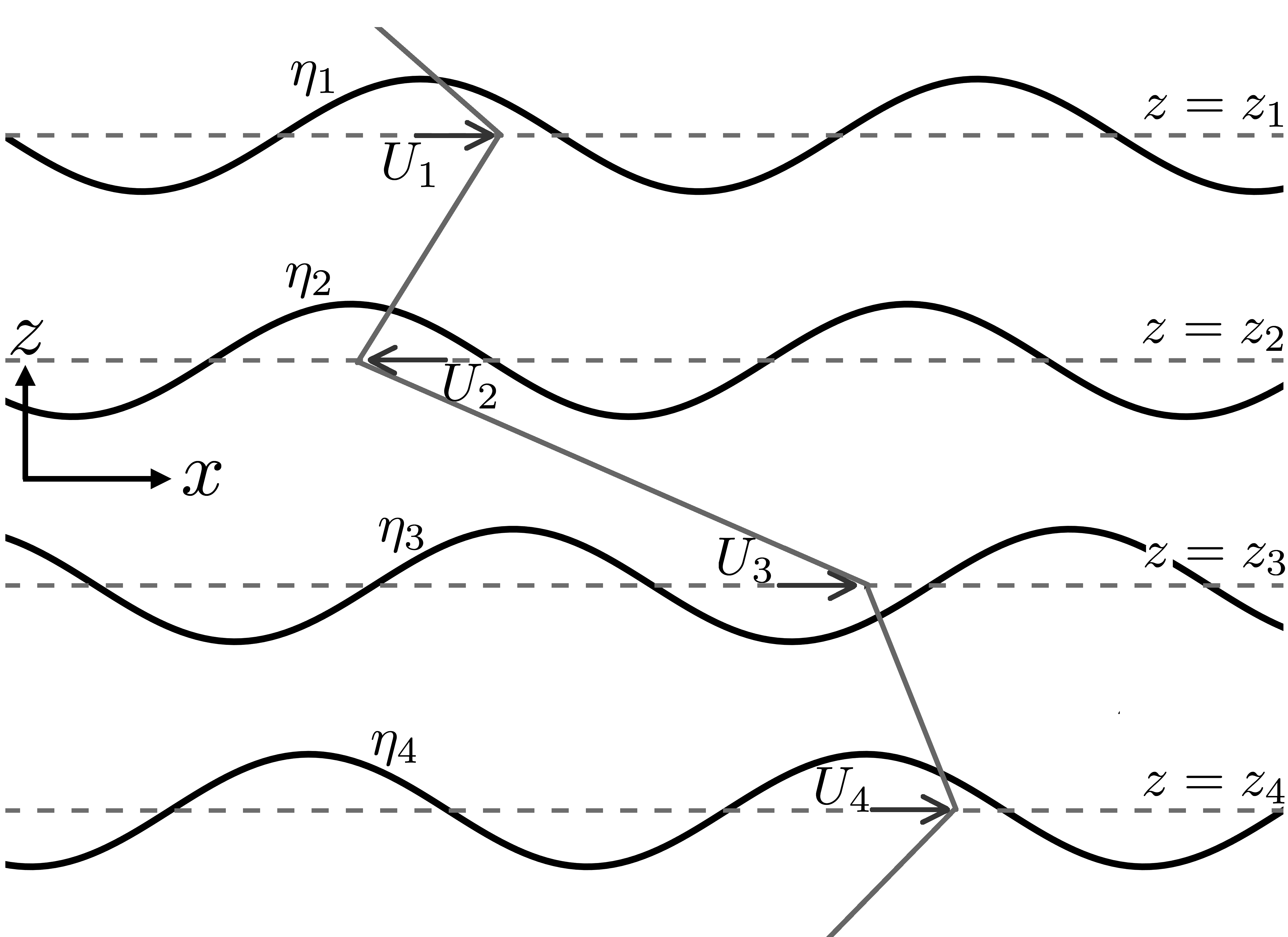}
}
\centering
~~~~\subfloat[]{\includegraphics[width=1.8in]{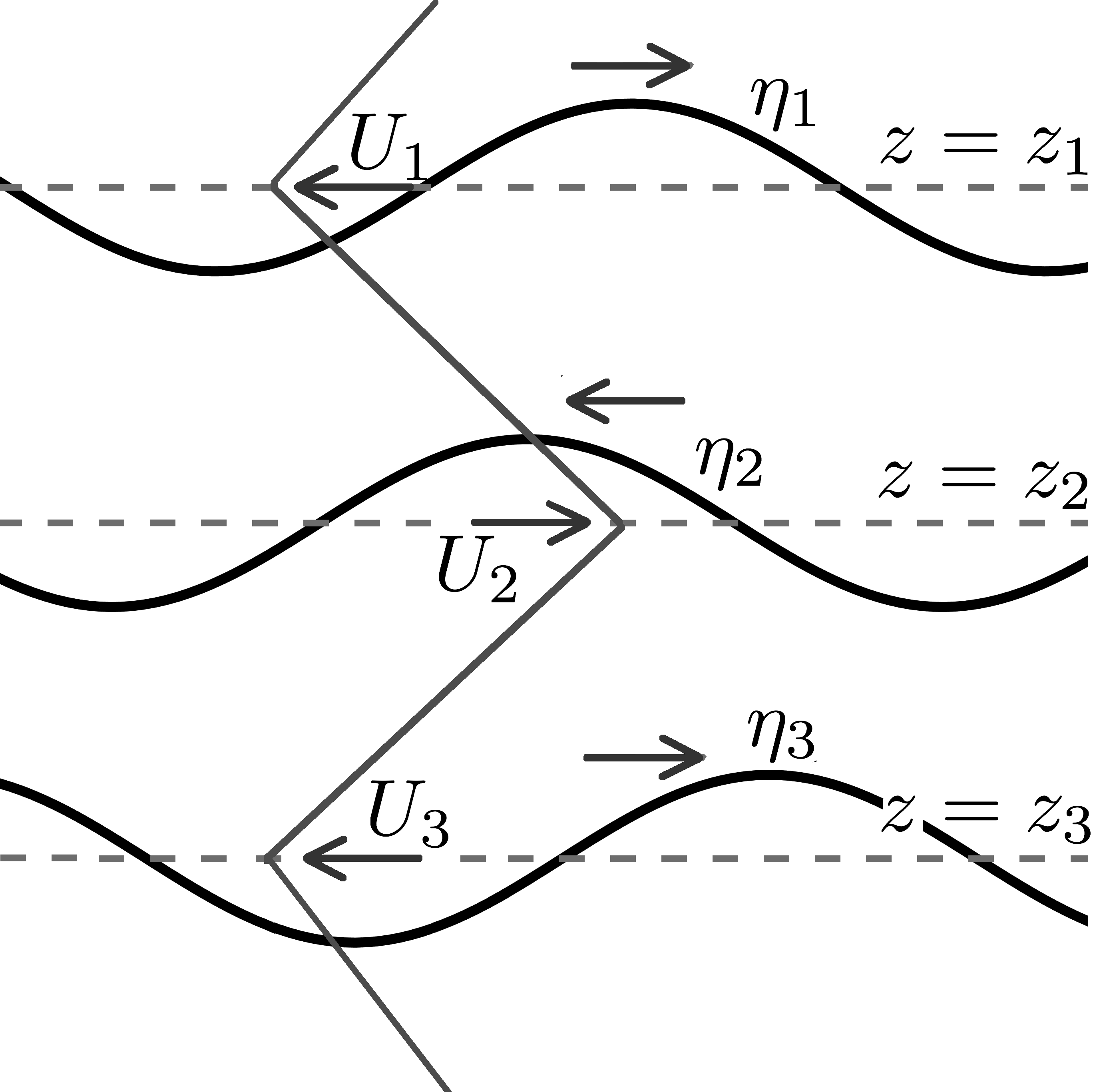}
}
\caption{(a) Schematic of the general set-up. (b)  The saw-tooth jet profile, which produces a $3$-interface problem with symmetry.}
\label{fig:schematic_basic}
\end{figure}







\section{The General Model}
We consider an inviscid, incompressible, 2D flow with $M$ interfaces, which are located at $z=z_{1}, z_{2},\dots z_{M}$ (see figure \ref{fig:schematic_basic}(a)).
 The last/boundary interfaces could be followed by an infinite medium.
The background velocity  $U$ is  parallel to the $x$ axis and is a piece-wise continuous function of $z$. Density may be constant or variable; in the latter case it is assumed to be layer-wise constant and decreasing with the vertically upward pointing coordinate $z$, implying stable stratification. When sinusoidal streamwise perturbations are added to such a layered fluid system, the resultant wave field becomes such that the waves propagate only along the interfaces \cite[]{suth2010}. The generation mechanism of this wave field can be described by the Poisson equation relating the perturbation stream-function $\psi(x,z,t)$ and the perturbation vorticity $q(x,z,t)$ \cite[]{draz1982,suth2010}: 
 \begin{equation}
\nabla^{2}\psi=q.
\label{eq:second}
\end{equation}
 We assume  $\psi$ and $q$ to represent sinusoidal disturbances along the $x$ direction. Furthermore, the disturbances are monochromatic with a wavenumber $\alpha$. This allows us to apply the Fourier ansatzs $q=\Re{\{ \hat{q}(z,t)\ee^{\ii\alpha x} \}}$ and $\psi=\Re{\{ \hat{\psi}(z,t)\ee^{\ii\alpha x} \}}$:
\begin{equation}
\left(\frac{\partial ^{2}}{\partial z^{2}}-\alpha^{2}\right)\psi=q. 
\label{eq:third}
\end{equation}
The above equation is a regular, non-homogeneous Sturm-Liouville problem with homogeneous boundary conditions: $\psi \rightarrow 0$ as $z \rightarrow \pm \infty$. It can be solved by inverting the linear operator on the left hand side of (\ref{eq:third}), yielding
\begin{equation}
 \psi=\intop_{\mathscr{B}}\mathscr{G}(s,z; \alpha) \,q \,ds,
\label{eq:fourth}
\end{equation}
where $\mathscr{B}$ is the field domain and $\mathscr{G}(s,z;\alpha)$ is the  Green's function satisfying $\partial^{2}\mathscr{G}/\partial z^{2}-\alpha^{2}\mathscr{G}=\delta(z-s)$, with the appropriate  boundary conditions. 
Our domain is unbounded (extending to $\pm \infty$),  which yields $\mathscr{G}=-\ee^{-\alpha\left|z-s\right|}/(2\alpha)$. 

In  inviscid flows, a particle on an interface $\eta_{j}=\eta(x,z_{j},t)$  stays on that interface forever. This is expressed in terms of the kinematic condition:
\begin{equation}
\frac{D\eta_{j}}{Dt} \equiv \frac{\partial\eta_{j}}{\partial t}+U_{j}\frac{\partial\eta_{j}}{\partial x}=w_{j}.
\label{eq:kinematic1}
\end{equation}
The  above equation is  the ``linearized'' kinematic condition (hence $D/Dt$ is the linearized material derivative operator) because the background velocity, $U_{j} \equiv U(z_{j})$, is known. $U_{j}$ should not be confused with  the 
perturbation $x$-velocity at the $j$-th interface, which is $u_{j} = \partial\psi_{j}/\partial z$. The quantity $w_{j}=w(x,z_{j},t)$ is the $z$-velocity at the $j$-th interface. Noting that $w=-\partial\psi/\partial x=-\ii \alpha \psi$, 
the linearized kinematic condition at the $j$-th interface can be expressed in terms of (\ref{eq:fourth}) as
\begin{equation}
\frac{D\eta_{j}}{Dt} = \frac{\ii}{2}\intop_{\mathscr{B}}\ee^{-\alpha\left|z_{j}-s\right|}\, q\, ds.\label{eq:kinematic2}
\end{equation}
Until this point we have only worked with different kinematic equations. Dynamics can enter into the problem through the $q$ term.
In 2D, inviscid, Boussinesq flows, the linearized perturbation vorticity evolution equation reads \cite[]{rabinovich2011,carp2012}
\begin{equation}
\frac{D q}{D t}=-w\frac{dQ}{dz}+N^{2}\frac{\partial\eta}{\partial x},
\label{eq:vort_evol}
\end{equation}
where  $Q=dU/dz$ is the background vorticity and $N(z)=\sqrt{-(g/\rho_{0}) d\bar{\rho}/dz}$ is the  background buoyancy frequency ($\rho_{0}$ is the reference density,  $\bar{\rho}$ is the background density,  and $g$ is gravity). The first term on the right hand side of (\ref{eq:vort_evol}) is known as the barotropic generation of vorticity (which is a kinematic process), while the second term implies baroclinic generation (which is a dynamic process). There are even other ways of dynamic generation of vorticity, e.g.\ magnetic fields \cite[]{heifetz2015},  surface tension \cite[]{biancofiore2015}, etc.

As an example we consider the simplest case where the flow is homogeneous/barotropic, i.e.\ we set $N=0$ in (\ref{eq:vort_evol}). Along with this equation we use the linearized kinematic condition $D\eta/Dt=w$, yielding
\begin{equation}
\frac{D q}{D t}=-\frac{D}{D t}\left(\eta \frac{ dQ}{dz}\right)\,\,\,\mathrm{which\,\,implies}\,\,\, q=-\eta\frac{dQ}{dz}.
\label{eq:homo_q}
\end{equation}  
In flows where the background vorticity $Q$ is layered,  one can  approximate $Q$ by a piece-wise constant function.
 This leads to a  considerable analytical simplification because the quantity $dQ/dz$ yields delta functions at each isolated discontinuity $z=z_{1},\,z_{2,} \cdots ,\, z_{M}$:
\begin{equation}
\frac{dQ}{dz}=\sum_{j=1}^{M}\Delta Q_{j}\delta\left(z-z_{j}\right).
\label{eq:delta_behav}
\end{equation}
Here  $\Delta Q_{j}\equiv Q({z^{+}_{j}})-Q({z^{-}_{j}})$ is the jump  in $Q$ at the discontinuity $z_{j}$. Equation (\ref{eq:delta_behav}) is substituted in (\ref{eq:homo_q}), and then the resultant expression is substituted in (\ref{eq:kinematic2}) to yield
 \begin{equation}
\frac{D\eta_{j}}{Dt}=\sum_{k=1}^{M}\tilde{w}_{k}(x,t)\thinspace e^{-\alpha z_{jk}},\label{eq:kinematic}
\end{equation}
where  $\tilde{w}_{k}=-\ii \eta_{k}\Delta Q_{k}/2$
and $z_{jk}=|z_{j}-z_{k}|$. We note here that $w_{j}$ of (\ref{eq:kinematic1}) has been expressed in (\ref{eq:kinematic})  as the sum of $z$-velocity contributions from all the $M$ interfaces, including itself.

In order to convert (\ref{eq:kinematic}) into a system of ODEs, we will assume Fourier ansatzs (and \emph{not} the conventional normal-mode ansatzs): 
$\eta_{j}(x,t)=\Re\{{A_{\eta_{j}}(t)e^{\ii(\alpha x+\phi_{\eta_{j}}(t))}\}} $ and $\tilde{w}_{j}(x,t)=\Re{\{A_{w_{j}}(t)e^{\ii(\alpha x+\phi_{w_{j}}(t))}\}}$,
where $A_{\eta_{j}}$, $A_{w_{j}}$, $\phi_{\eta_{j}}$ and $\phi_{w_{j}}$ are arbitrary real functions of $t$. We define the amplitude ratios
$\Omega_{j} \equiv A_{w_{j}}/A_{\eta_{j}}$ and $R_{jk} \equiv A_{\eta_{k}}/A_{\eta_{j}}$, and the phase differences $\Phi_{jk} \equiv \phi_{w_{k}}-\phi_{\eta_{j}}$. These definitions lead to the following identities, which will be used in the equations appearing later on in the article: 
\begin{align}
 & (i)\,R_{jk}=1/R_{kj},  (ii)\,R_{jk}=R_{jl}\cdot R_{lk},  (iii)\,\Phi_{jk}=\Phi_{kk}+\phi_{\eta_{k}}-\phi_{\eta_{j}}, \nonumber \\
 & (iv)\,\Phi_{jk}=\Phi_{jj}+\Phi_{kk}-\Phi_{kj},(v)\,\Phi_{jk}=\Phi_{jl}+\Phi_{lk}-\Phi_{ll},(vi)\,\Phi_{jk}=\Phi_{kk}+\Phi_{jl}-\Phi_{kl}.\label{eq:iden}
\end{align}
The above mentioned variables have the following range of values: $R_{jk} \in (0,\infty)$, $\Omega_{j} \in (0,\infty)$ and $\Phi_{jk} \in [-\pi,\pi]$,  where $j$ and $k$ are $1,2,..
M$. Waves whose intrinsic propagation is leftward have $\Phi_{jj}=\pi/2$, while those propagating rightward have $\Phi_{jj}=-\pi/2$ (the reason is explained below). 
Substitution of the Fourier ansatzs for $\eta_{j}$ and $\tilde{w}_{j}$  in (\ref{eq:kinematic}) produces
\begin{align}
 &   \dot{A}_{\eta_{j}}=\sum_{k=1}^{M}\Omega_{k}A_{\eta_{k}}\cos\left(\Phi_{jk}\right)e^{-\alpha z_{jk}},\label{eq:11}\\
 &   \dot{\phi}_{\eta_{j}}=-\alpha U_{j}+\sum_{k=1}^{M}\Omega_{k}R_{jk}\sin\left(\Phi_{jk}\right)e^{-\alpha z_{jk}}, \label{eq:12}
\end{align}
where $j=1,2,..
M$. While $\dot{A}_{\eta_{j}}$ in  (\ref{eq:11}) is the rate of change of wave amplitude,  $-\dot{\phi}_{\eta_{j}}$ in (\ref{eq:12}) implies the wave frequency. $\Omega_{k}$ has the dimensions of frequency, and is in fact the magnitude of the \emph{intrinsic frequency} of an interfacial wave in isolation. This can be shown as follows.  Consider a system with a single interface, i.e.\ $M=1$  in  (\ref{eq:11})-(\ref{eq:12}).  Since a wave cannot grow on its own, we must have $\dot{A}_{\eta}=0$ (index dropped for convenience), thereby implying 
 $\Phi=\pm \pi/2$. In  (\ref{eq:12}) $M=1$ also implies  $R=1$, hence this equation becomes $\dot{\phi}_{\eta}=-\alpha U\pm \Omega$. In the absence of background velocity/Doppler shift we have  $\dot{\phi}_{\eta}=\pm \Omega$, hence $\Omega$ is indeed the intrinsic frequency of an interfacial wave in isolation. The positive and negative signs respectively implying left and right moving waves. Usually the value of $\Omega$ comes from the dynamics, and is obtained from the dispersion relation $\mathcal{D}(\Omega,\alpha)=0$. For example, $\Omega$ of a long interfacial wave existing at the interface of two  fluid layers  of different densities (layer thicknesses respectively being $h_{1}$ and $h_{2}$) under the Boussinesq approximation is  $\alpha [g'h_{1}h_{2}/(h_{1}+h_{2})]^{1/2}$ \cite[]{suth2010}, where $g'$ is the reduced gravity.

It is convenient to define growth-rate $\sigma_{j}$ of the j-th interfacial wave as follows:
\begin{equation}
\sigma_{j}\equiv \dot{A}_{\eta_{j}}/A_{\eta_{j}}=\sum_{k=1}^{M}\Omega_{k}R_{jk}\cos\left(\Phi_{jk}\right)e^{-\alpha z_{jk}}.
\label{eq:grw_rate}
\end{equation}
Equations \ (\ref{eq:11})-(\ref{eq:12}) or (\ref{eq:12})-(\ref{eq:grw_rate}) emphasize the fact that the growth-rate $\sigma_{j}$ and frequency  $-\dot{\phi}_{\eta_{j}}$ of a wave at the $j$-th interface are governed by the linear interaction of all interfacial waves present in the system. Moreover the interaction model  (\ref{eq:11})-(\ref{eq:12}) is essentially \emph{kinematic}, the physics or dynamics are contained \emph{only} in the $\Omega_{k}$ terms. The advantage of being \emph{physics independent} is that the model is applicable to a wide variety of problems.

It is convenient to recast (\ref{eq:11})-(\ref{eq:12}) in terms of $R_{jk}$ and $\Phi_{jk}$:
\begin{align}
   & \dot{R}_{jk}=R_{jk}\sum_{l=1}^{M}\Omega_{l}\{ R_{kl}\cos\left(\Phi_{kl}\right)e^{-\alpha z_{kl}}-R_{jl}\cos\left(\Phi_{jl}\right)e^{-\alpha z_{jl}}\},\label{eq:13}\\
   & \dot{\Phi}_{jk}=\alpha\left(U_{j}-U_{k}\right)+\sum_{l=1}^{M}\Omega_{l}\{ R_{kl}\sin\left(\Phi_{kl}\right)e^{-\alpha z_{kl}}   -R_{jl}\sin\left(\Phi_{jl}\right)e^{-\alpha z_{jl}} \},\label{eq:14}
\end{align}
where both $j$ and $k$ are $1,2,..
M$. The above equation-set represents an autonomous, \emph{non-linear} dynamical system in $R_{jk}$ and $\Phi_{jk}$. Following the convention of GL14 we will refer to the model given by  (\ref{eq:13})-(\ref{eq:14}) as WIT. While the WIT equations of GL14 (their (3.9)-(3.12)) are limited to two interfaces, here we have generalized the problem to $M$ interfaces.
 We should note the apparently surprising non-linearity in the WIT equations  given that they are derived from (\ref{eq:kinematic}), which is a linear PDE. It should be further noted that the WIT equations of GL14, being the $2$-interface version of (\ref{eq:13})-(\ref{eq:14}), are  also non-linear. However, the phase space of GL14 is limited in its richness  since the dynamical system is only 2D.
 
 The fixed points of (\ref{eq:13})-(\ref{eq:14})  are of particular interest. In (\ref{eq:14}), the condition $\dot{\Phi}_{jk}=0$ implies $\dot{\phi}_{\eta_{j}}=\dot{\phi}_{\eta_{k}}$ (by using identity (iii) of  (\ref{eq:iden})), which means \emph{phase-locking} of the  waves located at the $j$-th and $k$-th interfaces. Furthermore, if $\sigma_{j}=\sigma_{k}=\mathrm{constant}$, the amplitudes of all the waves present in the system will have exponential growth or decay at the same rate. The condition $\dot{R}_{jk}=0$ in  (\ref{eq:13}) implies $\sigma_{j}=\sigma_{k}$, since 
  $\dot{R}_{jk}=R_{jk}(\sigma_{k}-\sigma_{j})$.
  This is the growth-rate that would have been obtained if the normal-mode ansatz were substituted in  (\ref{eq:kinematic}), instead of the Fourier ansatz.  The fixed points denote amplitude and phase locking, a state which we will refer to  as  \emph{wave synchronization} (while looking at it from the WIT perspective). This state, when looked at from the view-point of canonical linear stability theory, will be the normal-modes of the system.
 \section{ The 3-Interface (Saw-tooth jet) Problem}
 We investigate WIT for a system that has three interfaces and an inherent kinematic and geometric symmetry. For this we have chosen the saw-tooth jet flow profile, see figure~\ref{fig:schematic_basic}(b). It approximates the multiple zonal-jet flow structure in planetary atmospheres resulting from potential vorticity staircases \cite[]{drit2008,scott2012}.  In this system 
$\Omega_{1}=\Omega_{2}=\Omega_{3}=\Omega$, $U_{1}=U_{3}$ and $z_{12}=z_{32}=Z$. Our set-up is different from the triangular jet profile \cite[]{draz1982}, where $\Omega_{1}=\Omega_{3}=\Omega$ and $\Omega_{2}=2\Omega$ (note that our analysis will hold for triangular jet profile also). We use the non-dimensional time $\tau=\Omega t$, and hereafter denote $\dot{\left\langle \,\, \right\rangle} \equiv d/d\tau$. A ``Froude number'' like dimensionless variable is defined by $\gamma \equiv \alpha(U_{2}-U_{1})/\Omega$. Without any loss of generality, $U_{1}$ and $U_{3}$ are taken as $0$, and $U_{2}\geq 0$, which implies $\gamma \geq 0$. The interfacial waves are assumed to ``counter-propagate'', i.e. travel in a direction opposite to the background flow at that interface. Hence the intrinsic propagation of wave-$2$ is leftward (i.e. $\Phi_{22}=\pi/2$).  Wave-$1$ and wave-$3$ have intrinsic propagation to the right (i.e. $\Phi_{11}=\Phi_{33}=-\pi/2$). The wave amplitudes and phases evolve as follows: 
\begin{align}
 & \dot{A}_{\eta_{1}}=A_{\eta_{2}}\cos\left(\Phi_{12}\right)e^{-\alpha Z}+A_{\eta_{3}}\sin\left(\Phi_{12}-\Phi_{32}\right)e^{-2\alpha Z},\label{eq:121}\\
 & \dot{A}_{\eta_{2}}=A_{\eta_{1}}\cos\left(\Phi_{12}\right)e^{-\alpha Z}+A_{\eta_{3}}\cos\left(\Phi_{32}\right)e^{-\alpha Z},\label{eq:122}\\
 & \dot{A}_{\eta_{3}}=A_{\eta_{1}}\sin\left(\Phi_{32}-\Phi_{12}\right)e^{-2\alpha Z}+A_{\eta_{2}}\cos\left(\Phi_{32}\right)e^{-\alpha Z},\label{eq:123}\\
  & \dot{\Phi}_{12}=-\gamma+2- e^{-\alpha Z}[(\frac{A_{\eta_{1}}}{A_{\eta_{2}}}+\frac{A_{\eta_{2}}}{A_{\eta_{1}}})\sin\left(\Phi_{12}\right)+\frac{A_{\eta_{3}}}{A_{\eta_{2}}}\sin\left(\Phi_{32}\right)-\frac{A_{\eta_{3}}}{A_{\eta_{1}}}\cos\left(\Phi_{12}-\Phi_{32}\right)e^{-\alpha Z}], \label{eq:124}\\
 & \dot{\Phi}_{32}=-\gamma+2- e^{-\alpha Z}[(\frac{A_{\eta_{2}}}{A_{\eta_{3}}}+\frac{A_{\eta_{3}}}{A_{\eta_{2}}})\sin\left(\Phi_{32}\right)+\frac{A_{\eta_{1}}}{A_{\eta_{2}}}\sin\left(\Phi_{12}\right)-\frac{A_{\eta_{1}}}{A_{\eta_{3}}}\cos\left(\Phi_{12}-\Phi_{32}\right)e^{-\alpha Z}]\label{eq:125}.
\end{align}
These equations have been simply obtained by applying the saw-tooth jet setting to  (\ref{eq:11}) and using (\ref{eq:iden}). We observe similarity between (\ref{eq:121})-(\ref{eq:123}) and the amplitude evolution equations of the triangular jet problem  studied by \cite{heif1999} (see their equations (19a)-(19c)). Manipulation of (\ref{eq:121})-(\ref{eq:123}) yields a conservation equation
\begin{equation}
A_{\eta_{1}}^{2}+A_{\eta_{3}}^{2}-A_{\eta_{2}}^{2}=\textrm{constant}.
\label{eq:energy}
\end{equation}
We found that  exactly the same conservation equation can be obtained for a triangular jet. Finding a conservation equation for perturbation quantities in the presence of a background flow is not usually possible, and the common approach is to find a conserved wave activity (or activities). Equation (\ref{eq:energy}) seems to be a special case in this regard.

\begin{figure}
\centering
\includegraphics[width=5.3in]{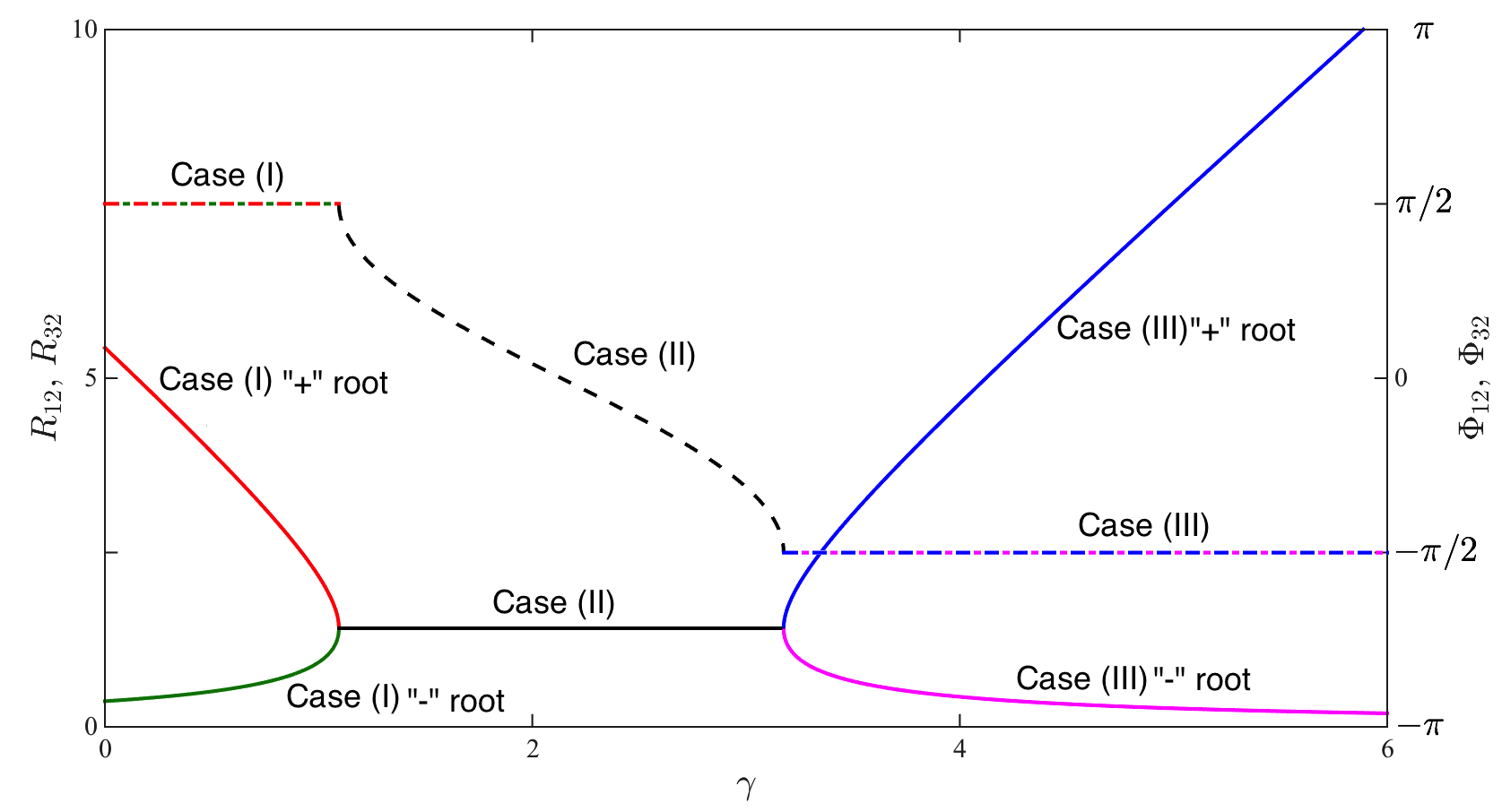}
\caption{ (Color online) Variation of fixed points with $\gamma$ for $\alpha=1$ and $Z=1$. Solid lines indicate $R_{12}$ and $R_{32}$,while dashed lines indicate $\Phi_{12}$ and $\Phi_{32}$.}
\label{fig:fixed_vs_gamma}
\end{figure}

We recast (\ref{eq:121})-(\ref{eq:125}) in terms of  amplitude ratios. After some algebra and use of (\ref{eq:iden}),  the following set of equations are obtained:
\begin{align}
& \dot{R}_{12}= e^{-\alpha Z}[(1-R_{12}^{2})\cos\left(\Phi_{12}\right)+\frac{R_{12}}{R_{32}}\cos\left(\Phi_{32}\right)-\frac{R_{12}^{2}}{R_{32}}\sin\left(\Phi_{12}-\Phi_{32}\right)e^{-\alpha Z}],\label{eq:3_set_1} \\
 & \dot{R}_{32}= e^{-\alpha Z}[(1-R_{32}^{2})\cos\left(\Phi_{32}\right)+\frac{R_{32}}{R_{12}}\cos\left(\Phi_{12}\right)+\frac{R_{32}^{2}}{R_{12}}\sin\left(\Phi_{12}-\Phi_{32}\right)e^{-\alpha Z}],\label{eq:3_set_2} \\
 & \dot{\Phi}_{12}=-\gamma+2- e^{-\alpha Z}[\frac{(1+R_{12}^{2})}{R_{12}}\sin\left(\Phi_{12}\right)+\frac{1}{R_{32}}\sin\left(\Phi_{32}\right)-\frac{R_{12}}{R_{32}}\cos\left(\Phi_{12}-\Phi_{32}\right)e^{-\alpha Z}], \label{eq:3_set_3} \\
 & \dot{\Phi}_{32}=-\gamma+2- e^{-\alpha Z}[\frac{(1+R_{32}^{2})}{R_{32}}\sin\left(\Phi_{32}\right)+\frac{1}{R_{12}}\sin\left(\Phi_{12}\right)-\frac{R_{32}}{R_{12}}\cos\left(\Phi_{12}-\Phi_{32}\right)e^{-\alpha Z}]. \label{eq:3_set_4}
\end{align}
The  above equations are basically the WIT equations (\ref{eq:13})-(\ref{eq:14}) for saw-tooth jet profile.
 It is comparatively easier to find the fixed points of the 4D system (\ref{eq:3_set_1})-(\ref{eq:3_set_4}) than the 5D system (\ref{eq:121})-(\ref{eq:125}).  Depending on the ranges of $\gamma$, different fixed points of (\ref{eq:3_set_1})-(\ref{eq:3_set_4}) are obtained (this has also been shown in  figure~\ref{fig:fixed_vs_gamma}):

\subsection{Case (I):}
 $\gamma\leq e^{-2\alpha Z}+2-2\sqrt{2}e^{-\alpha Z}$:
\begin{align}
& R_{12}=R_{32}=\frac{1}{2}[e^{\alpha Z}\left(e^{-2\alpha Z}+2-\gamma\right)\pm \nonumber \sqrt{e^{2\alpha Z}\left(e^{-2\alpha Z}+2-\gamma\right)^{2}-8}]\\ 
& \mathrm{and}\,\, \Phi_{12}=\Phi_{32}=\frac{\pi}{2}.
\label{eq:casei}
\end{align}
 \subsection{Case (II):}
 $e^{-2\alpha Z}+2-2\sqrt{2}e^{-\alpha Z} \leq \gamma\leq e^{-2\alpha Z}+2+2\sqrt{2}e^{-\alpha Z}$:
\begin{align}
& R_{12}=R_{32}=\sqrt{2} \,\,\mathrm{and} \nonumber \\
&  \Phi_{12}=\Phi_{32}=\sin^{-1}\left[\frac{1}{2\sqrt{2}}\left\{ e^{-\alpha Z}-\left(\gamma-2\right)e^{\alpha Z}\right\}\right].
\label{eq:caseii}
\end{align}

 \subsection{Case (III):}
\begin{figure}
\centering
\subfloat[]{\includegraphics[width=2.5in]{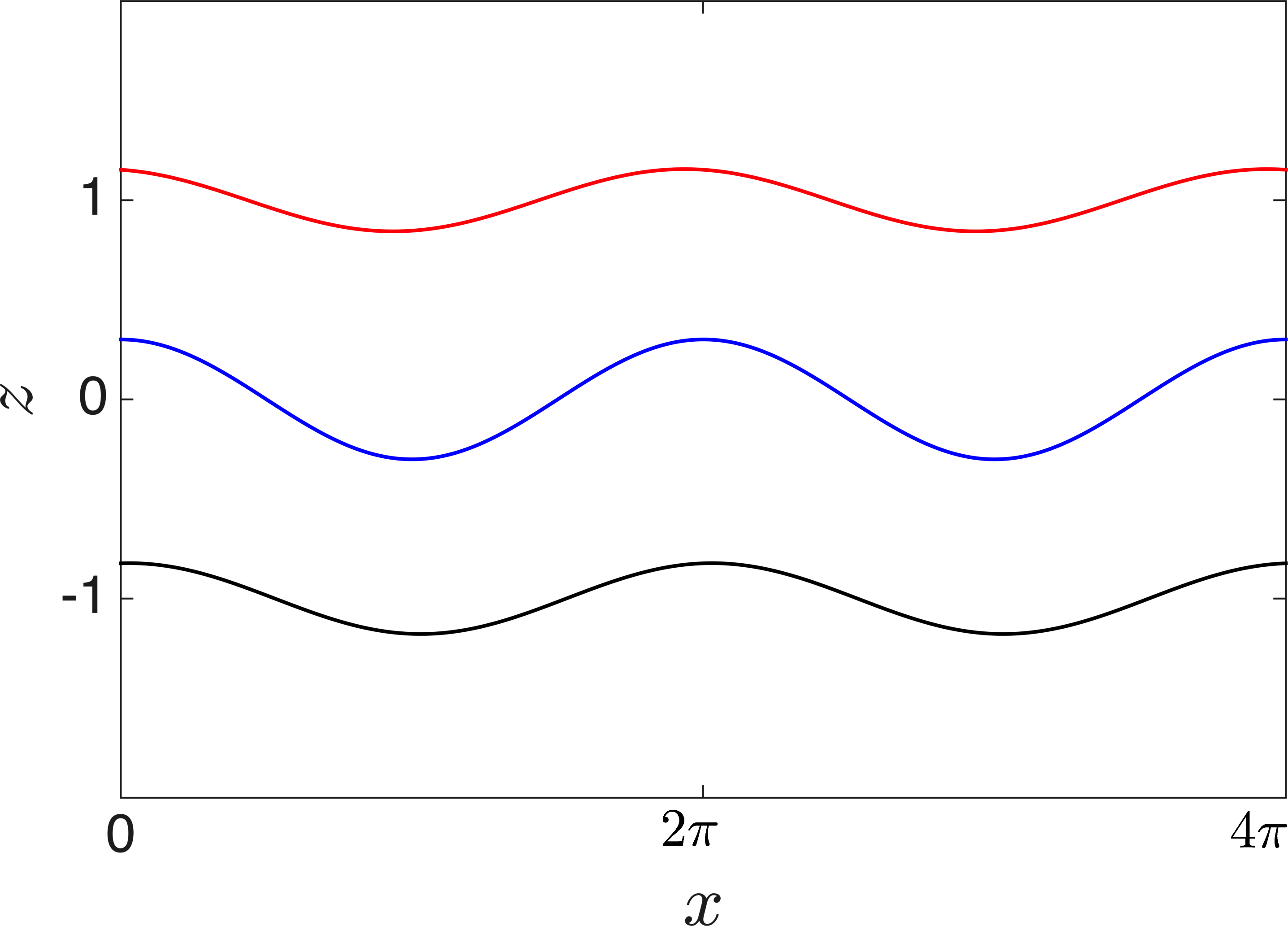}
}
\subfloat[]{\includegraphics[width=2.5in]{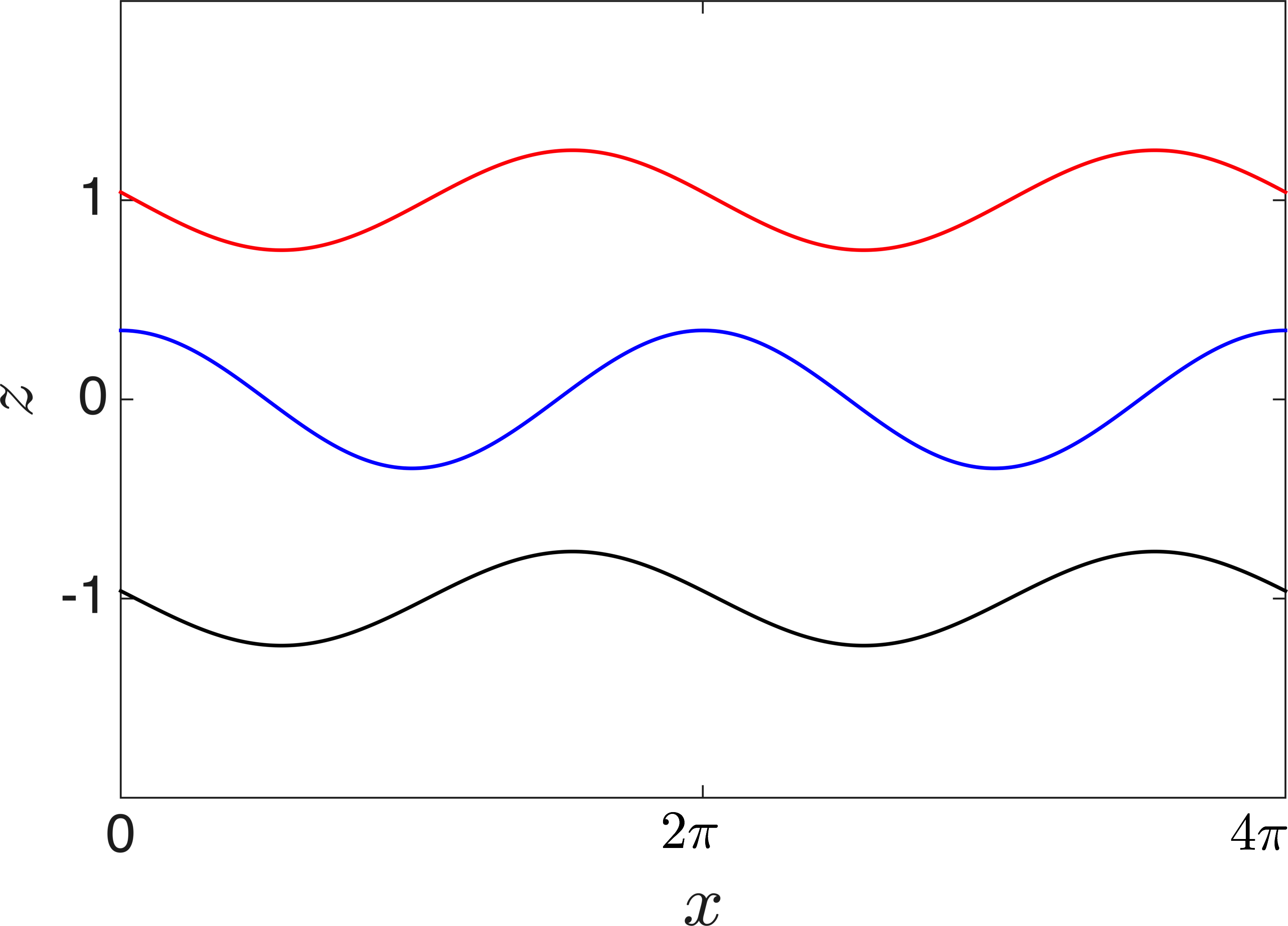}
}\\
\centering
\subfloat[]{\includegraphics[width=2.5in]{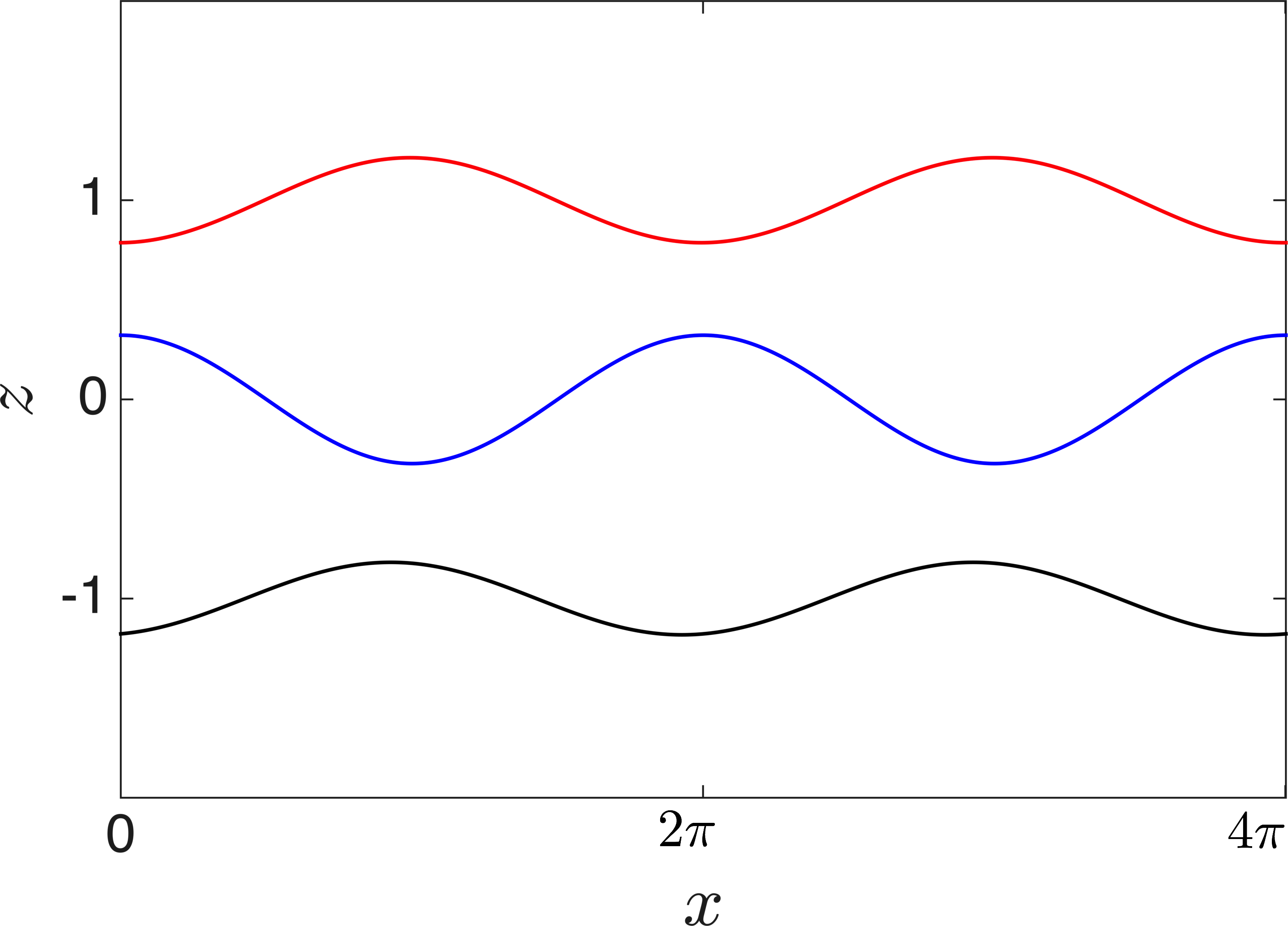}
}
\caption{  (Color online)  Three interfacial waves on reaching phase-locked configuration (fixed points). (a) $\gamma=1$, which corresponds to Case (I). Amplitudes have been exaggerated.  (b) $\gamma=2$, which corresponds to Case (II), which is an unstable normal-mode. (c) $\gamma=3.2$, which corresponds to Case (III). Amplitudes have been exaggerated.} 
\label{fig:amp_phase}
\end{figure}

  $\gamma\geq e^{-2\alpha Z}+2+2\sqrt{2}e^{-\alpha Z}$:
\begin{align}
& R_{12}=R_{32}=\frac{1}{2}[-e^{\alpha Z}\left(e^{-2\alpha Z}+2-\gamma\right)\pm  \sqrt{e^{2\alpha Z}\left(\gamma-2-e^{-2\alpha Z}\right)^{2}-8}]\nonumber\\ 
& \mathrm{and}\,\, \Phi_{12}=\Phi_{32}=-\frac{\pi}{2}.
\label{eq:caseiii}
\end{align}
 The derivations of Cases (I)-(III) are involved and are briefly outlined  in the Appendix. The fixed point configurations corresponding to each case are shown in figures \ref{fig:amp_phase}(a)-(c).  A pair of sinuous waves correspond to Case (I); see figure \ref{fig:amp_phase}(a). The phase-locked configuration of Case (II) is shown in figure \ref{fig:amp_phase}(b); it corresponds to normal-mode instability and the phase-shifts are dependent on $\gamma$; see (\ref{eq:caseii}).  Case (III) reveals a pair of varicose waves as shown in figure \ref{fig:amp_phase}(c). 

\begin{figure}
\centering
\subfloat[]{\includegraphics[width=2.5in]{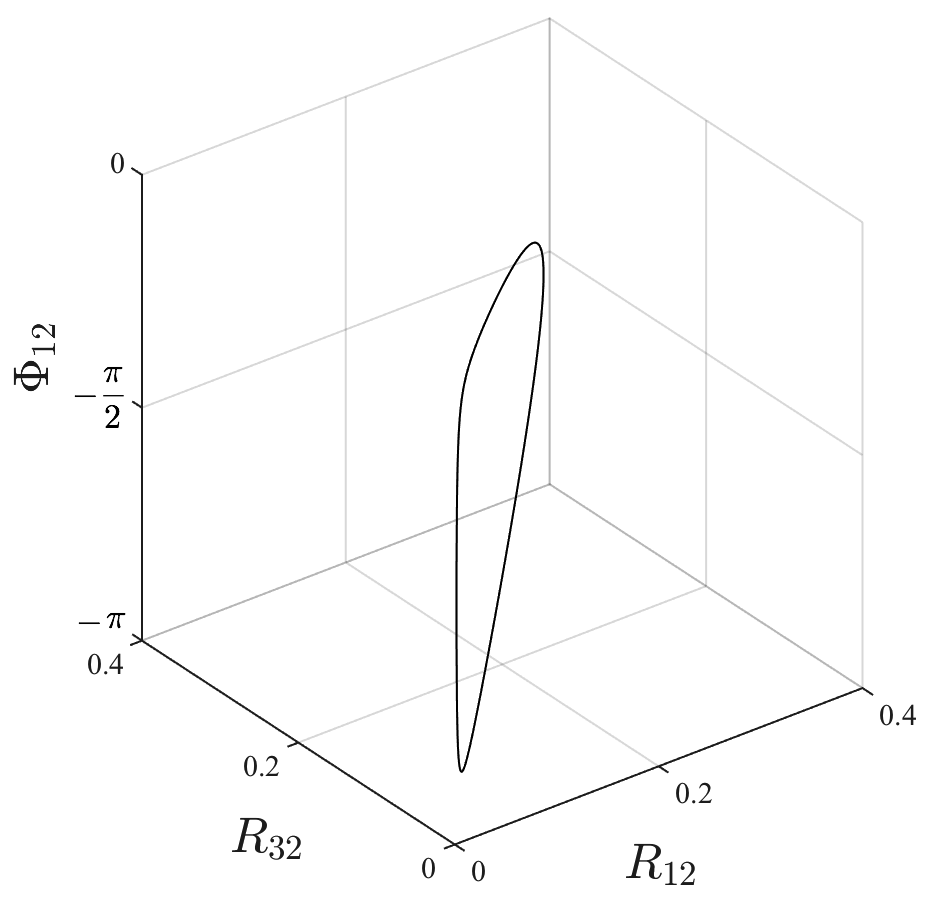}
}
\centering
\subfloat[]{\includegraphics[width=3.0in]{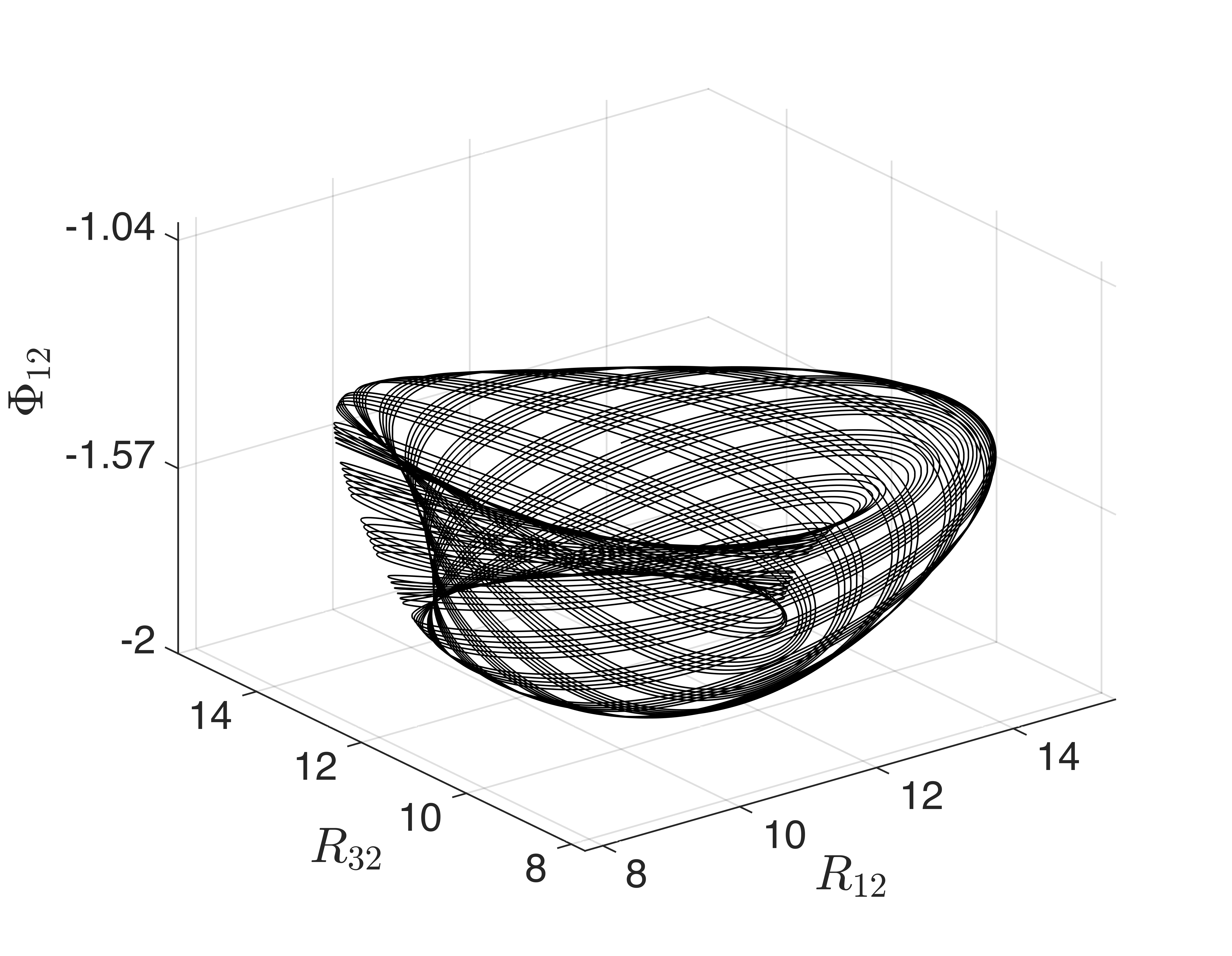}
}
\caption{ Behaviour around fixed points for $\gamma=6$ corresponding to different initial conditions: (a) initial condition $(R_{12},R_{32},\Phi_{12},\Phi_{32})=(0.01,0.01,-\pi/4,-\pi/4)$, and (b) initial condition $(R_{12},R_{32},\Phi_{12},\Phi_{32})=(10,15,-\pi/2,-\pi/2)$.}
\label{fig:orbit}
\end{figure}

In order to understand the nature of stability corresponding to each case mentioned above, we have computed the eigenvalues of the Jacobian matrix of the right hand side of  (\ref{eq:3_set_1})-(\ref{eq:3_set_4}) evaluated at the fixed points. In Case (II), all eigenvalues always have a negative real part, implying ``growing normal-mode'' (as shown in GL14). In other words, the range of $\gamma$ given in Case (II) allows normal-mode type instabilities.  Here wave synchronization is evident -  all the three waves are locked in amplitude and phase, and therefore grow at the same rate. The eigenvalues always have $0$ real part in Cases (I) and (III), and the fixed points appear to be unstable. Small perturbations from them lead to what appear to be periodic or pseudo-periodic orbits. As an example, for $\gamma=6$ (which corresponds to Case (III) when $\alpha=1$ and $Z=1$),  we found both periodic  and  pseudo-periodic orbits corresponding to different initial conditions, as shown in figure \ref{fig:orbit}.

 \begin{figure}
 \centering
\subfloat[]{\includegraphics[width=2.64in]{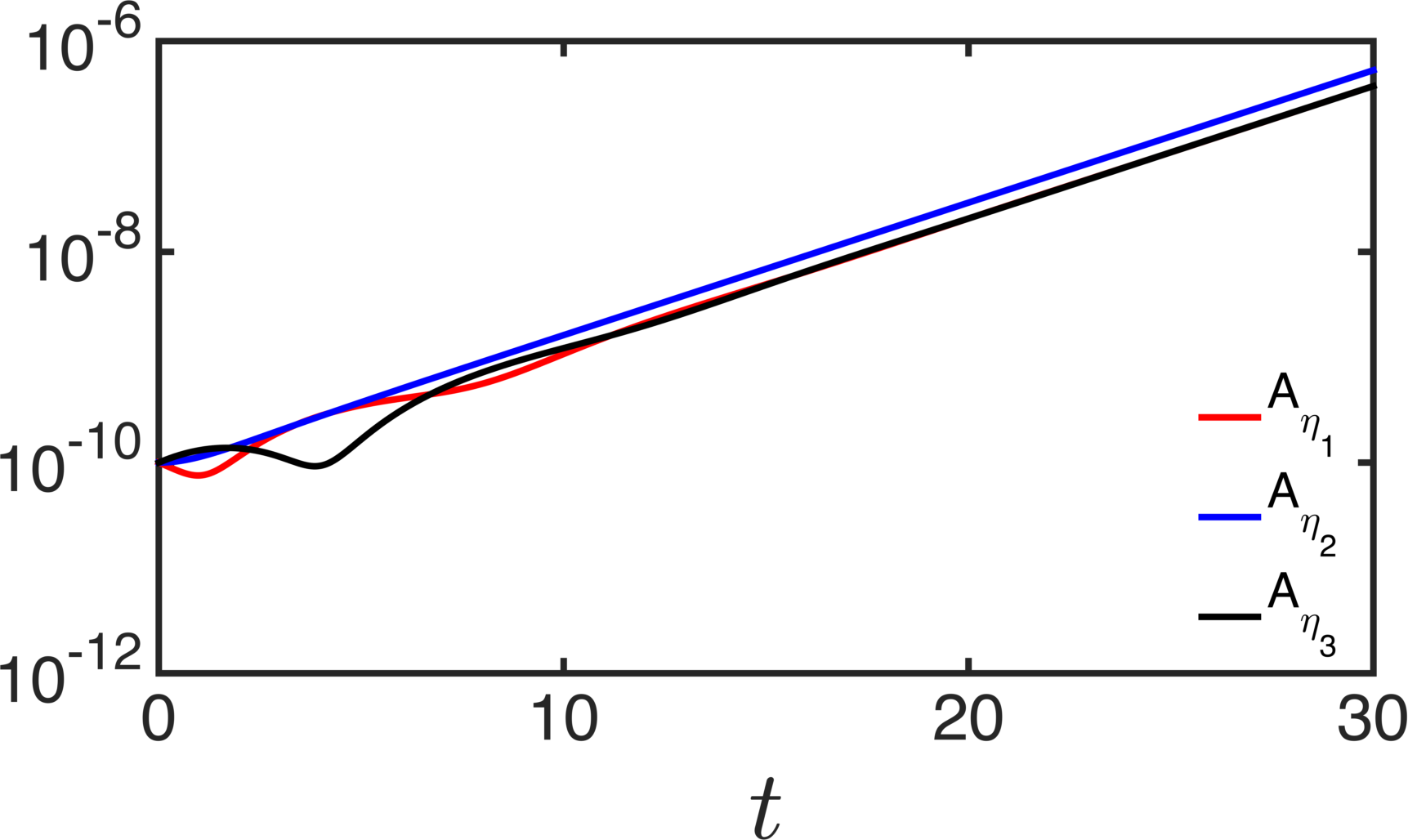}
}
\subfloat[]{\includegraphics[width=2.6in]{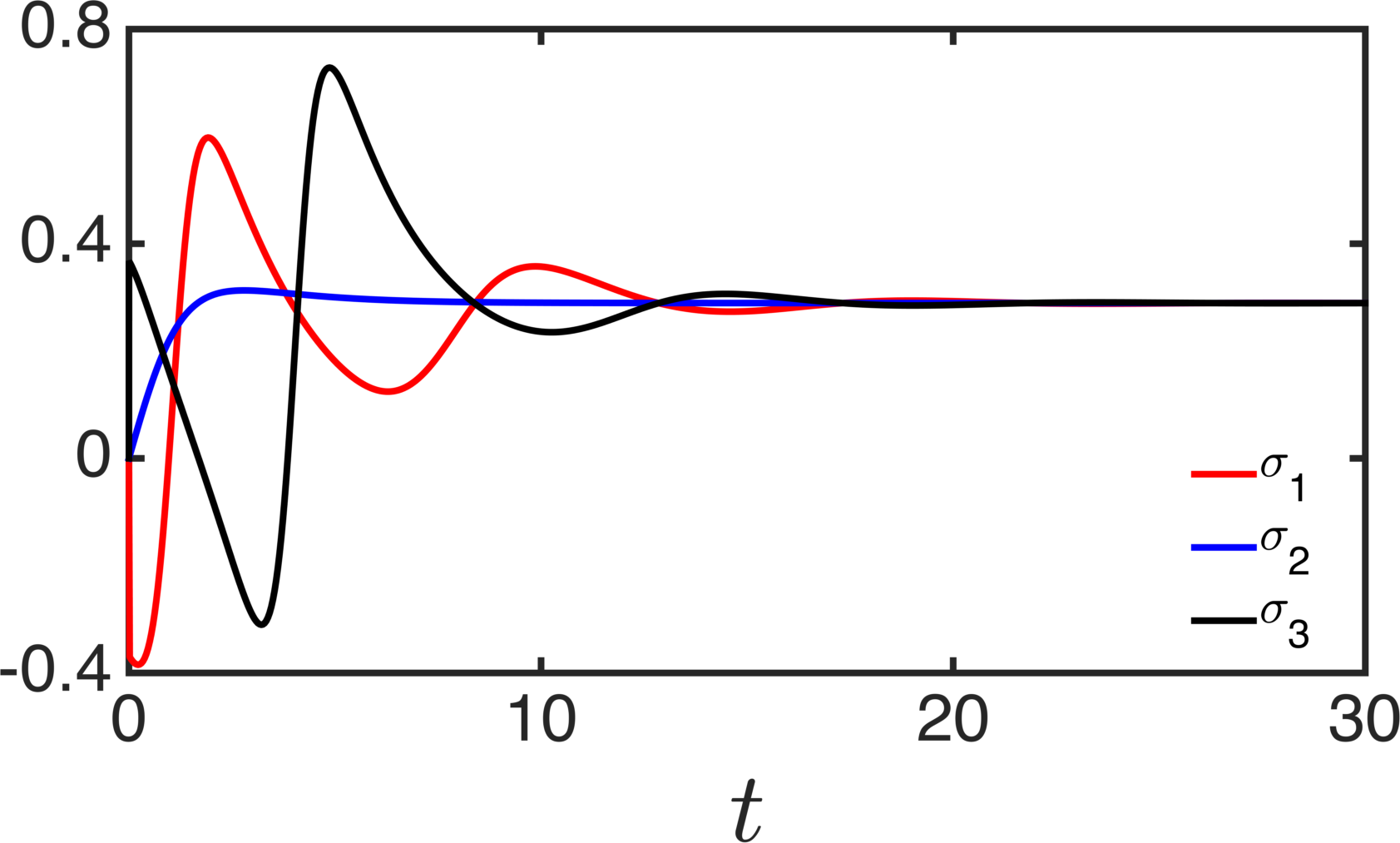}
}

 \centering
\subfloat[]{\includegraphics[width=2.7in]{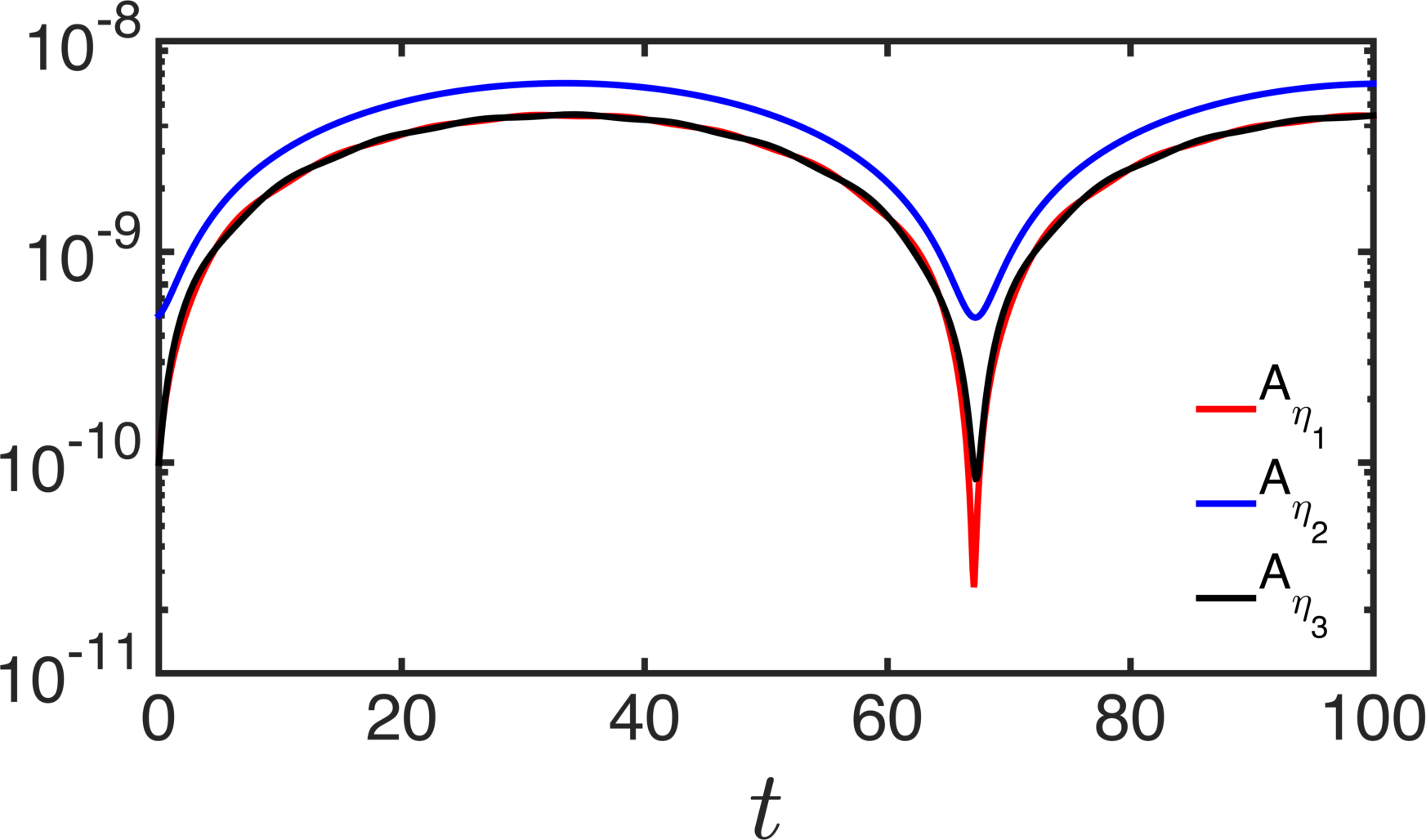}
}
\subfloat[]{\includegraphics[width=2.55in]{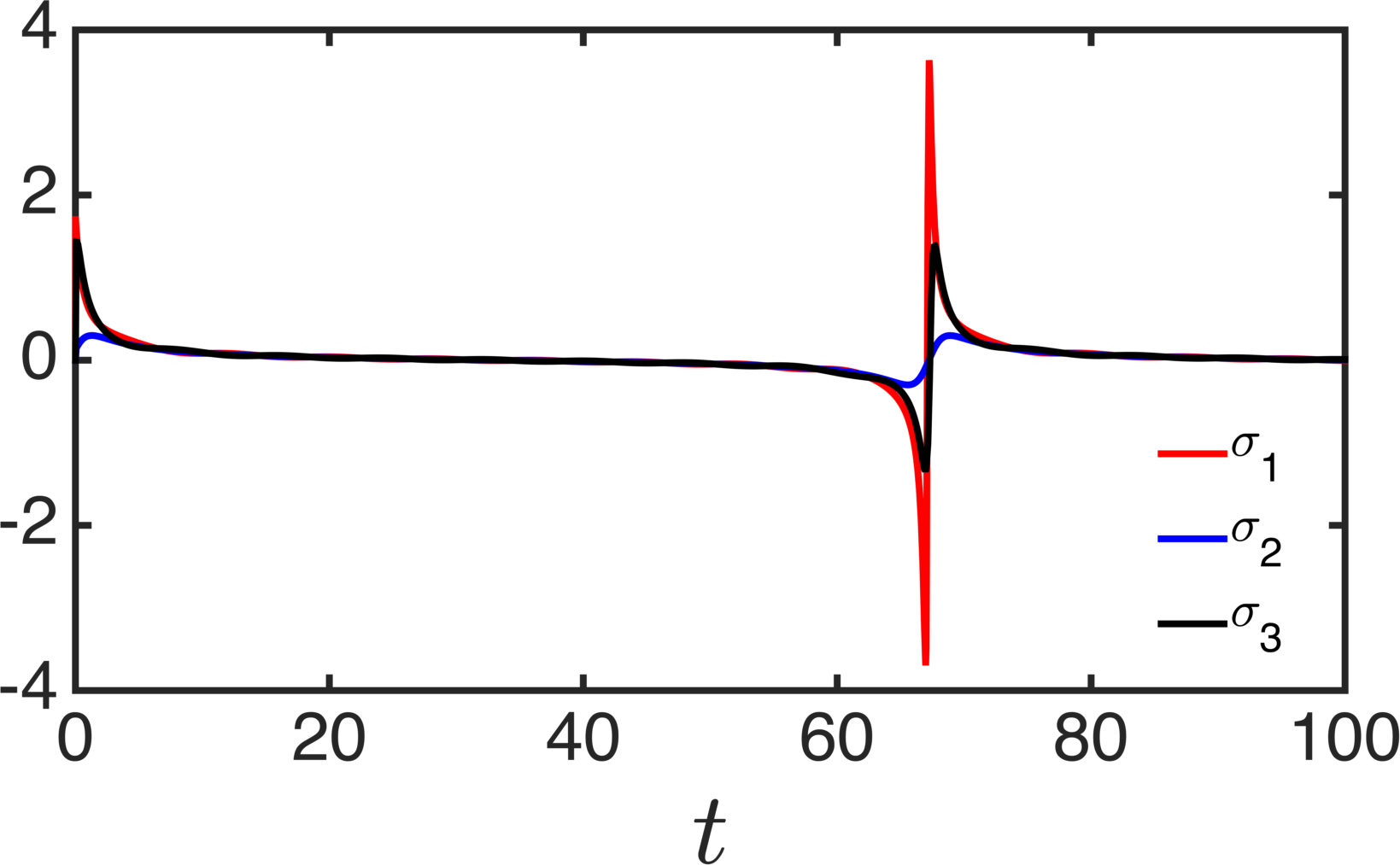}
}

\caption{ (Color online) Temporal variation of amplitudes and growth-rates of the constituent waves: (a) Amplitude and (b) growth-rate corresponding to $\gamma=3$ (Case (II)). (c) Amplitude and (d) growth-rate corresponding to $\gamma=3.18$ (Case (III)). }
\label{fig:gamma_vs_time}
\end{figure}

We also look at the temporal variation of amplitudes and growth-rates of each constituent wave. 
For normal-mode instability, all the waves have the same constant $\sigma$, which is only possible in Case (II) because there is only one root corresponding to $R_{12}$ or $R_{32}$. In figure~\ref{fig:gamma_vs_time}, we have plotted amplitude and growth-rate  of each  wave corresponding to $\gamma=3$ and $3.18$. In all our simulations $\alpha=1$ and $Z=1$. Hence $\gamma=3$ represents Case (II) while $\gamma=3.18$ represents Case (III), the latter representing behavior in the neighbourhood of the upper stability boundary ($\gamma = e^{-2\alpha Z}+2+2\sqrt{2}e^{-\alpha Z}$). Figures~\ref{fig:gamma_vs_time}(a)-(b) show that the constituent waves undergo transient growth/decay initially, but soon synchronize and attain the same normal-mode growth-rate. In this case the initial condition is $(1,1,-\pi,0)$. Case (III) (as well as Case (I)) represents that part of the parameter space for which canonical normal-mode theory would predict neutral stability.  As evident from figures \ref{fig:gamma_vs_time}(c)-(d), which correspond to  an initial condition of  $(5,5,0,\pi/4)$, transient growth/decay is possible.  In fact the waves grow more than an order of magnitude.   One or two orders of magnitude growth in amplitude may not be significant enough to introduce non-linearity in the system and alter the background flow through wave-mean interactions. However, large transient growth may arise in a more general setting shown in figure \ref{fig:schematic_basic}(a), and this hypothesis needs to be tested in future.


A significant aspect of WIT  is that it allows us to capture the transient dynamics of each wave separately. As shown in figure~\ref{fig:gamma_vs_time}(b), the three waves undergo different growth-rates initially. Such behavior cannot be properly captured using eigenvalue analysis (i.e. normal-mode stability theory), or SVD techniques (generalized stability theory) outlined in \cite{farrell1996}. While SVD analysis does capture transient growth, the growth-rate of all the constituent waves have to be the same. This growth-rate  is given by the maximum singular value, and  is known as  ``optimal growth'' in the literature. Clearly SVD analysis will not be able to predict the unequal growth/decay rates of the constituent waves during the initial period shown in figure \ref{fig:gamma_vs_time}(b). 

  We also have numerically calculated the Lyapunov exponents. Formally, the maximum Lyapunov exponent is defined as 
\[
\lambda_{max} \equiv \lim_{t\rightarrow\infty}\lim_{\delta\mathbf{X_{0}}\rightarrow0}\,\,\frac{1}{t}\ln\left(\frac{|\delta\mathbf{X(t)}|}{|\delta\mathbf{X_{0}}|}\right).
\]
It characterizes the exponential rate of separation of infinitesimally close trajectories whose initial separation is  $\delta\mathbf{X_{0}}$.  An autonomous non-linear dynamical system with  $\lambda_{max}>0$ is non-integrable, hence chaos is a possibility \cite[]{yoshida2010nonlinear}.  We have numerically computed the Lyapunov exponents up to $t= 10,000$ using the procedure outlined in \cite{wolf1985}. For $\gamma > 6$, the magnitudes of Lyapunov exponents oscillate between zero and a small positive number.  It is difficult to ascertain whether they will still remain positive even at very large times.

\section{Conclusions and Remarks}
WIT has previously been  studied to mainly provide a physical interpretation of shear instabilities. It turns out that in most situations, two interfacial waves are adequate in this regard. However, there are many geophysical flows where multiple interfacial waves are present, and analyzing their interactions is crucial for understanding those processes.  In this regard we have formulated a generalized theory to study interactions between $N$ linear interfacial waves. Moreover, the approach being kinematic (i.e.\ no need to specify the physics of the problem is advance), it is applicable to a wide range of physical problems.  By taking an expanded view of such interactions without making the commonly used normal-mode assumption, we have presented an apparently counter-intuitive phenomenon --  non-linear dynamics within the purview of linear wave theory. This phenomenon arises because the governing linear PDEs yield a non-linear autonomous dynamical system when Fourier ansatz is used instead of normal-mode anzatz. 

This general framework has been applied to a saw-tooth jet profile with three interfaces, yielding a 5D non-linear dynamical system (\ref{eq:121})-(\ref{eq:125}). 
 For a certain range of the Froude number like parameter $\gamma$, the system, usually after an initial transient growth or decay, gives rise to normal-mode instabilities. If one starts with the normal-mode ansatz at the outset, (s)he will find exponentially growing instabilities in this range of $\gamma$. Outside this range, normal-mode theory predicts neutral stability. However, in this apparently non-interesting range of $\gamma$, our more general Fourier ansatz formalism  shows that about an order of magnitude transient growth in amplitude is possible for some initial conditions. In more complicated systems it may so happen that one of the constituent waves can grow by many orders of magnitude, making the physical system non-linear.  The WIT framework enables the capture of different growth-rates of the constituent waves; eigen-analysis and  SVD would fail in this regard.

The range of $\gamma$ for which normal-mode theory predicts neutral stability is reasonably complex. The Lyapunov exponents calculated in this range (especially when $\gamma>6$) are found to oscillate between zero and a small positive number; it is difficult to ascertain whether they will remain so even at larger times.  In a later communication we therefore intend to further explore the possibility of chaos, especially when there are more than three interacting waves. Chaos may also appear when the kinematic and/or geometric symmetry of the 3-wave system are/is broken.
 

Outside the normal-mode parameter regime, non-linearity of the  5D dynamical system stemming from the  $3$-wave interaction problem gives rise to periodic and pseudo-periodic orbits in  phase space.  Fixed points bifurcate under small perturbations to exhibit periodic and pseudo-periodic behavior.  
This is in stark contrast to the 2D dynamical system stemming from the   $2$-wave interaction problem, which only reveals stable and unstable nodes in the normal-mode parameter range, and no fixed-points (or other interesting features) outside this range.



\section{Acknowledgment}
We are grateful to the anonymous referees for their useful comments and suggestions. AG thanks STC/ME/2016176 and PLANEX/PHY/2015239 for research support. 


\appendix
\section{ Derivation of the fixed points for the $3$-Interface Problem} 
 Fixed points of the system can be found by equating the right hand side of each of  (\ref{eq:3_set_1})-(\ref{eq:3_set_4}) to $0$. Subtracting (\ref{eq:3_set_1}) from (\ref{eq:3_set_2}) and imposing  $\dot{R}_{12}=0$ and $\dot{R}_{32}=0$, we obtain the following  conditions:
\begin{eqnarray}
 & & \mathrm{Either}\,\,\,\,\,\, R_{12}^{2}+R_{32}^{2}=R_{12}^{2}R_{32}^{2}   \,\,\,(\mathrm{Condition\,\, I}),        \\
& & \mathrm{Or}\,\,\,\,\,\, R_{12}\cos\left(\Phi_{32}\right)=-R_{32}\cos \left(\Phi_{12}\right)\,\,\, (\mathrm{Condition\,\, II}).
\end{eqnarray}
Furthermore, imposing $\dot{\Phi}_{12}=0$ and $\dot{\Phi}_{32}=0$ respectively in  (\ref{eq:3_set_3}) and (\ref{eq:3_set_4}), we obtain
\begin{eqnarray}
 &  & \left(\gamma-2\right)e^{\alpha Z}=-\frac{(1+R_{12}^{2})}{R_{12}}\sin\left(\Phi_{12}\right)-\frac{1}{R_{32}}\sin\left(\Phi_{32}\right)+\frac{R_{12}}{R_{32}}\cos\left(\Phi_{12}-\Phi_{32}\right)e^{-\alpha Z},\label{eq:4_set_1} \,\,\,\,\,\,\,\,\,\,\,\,\\
 &  & \left(\gamma-2\right)e^{\alpha Z}=-\frac{(1+R_{32}^{2})}{R_{32}}\sin\left(\Phi_{32}\right)-\frac{1}{R_{12}}\sin\left(\Phi_{12}\right)+\frac{R_{32}}{R_{12}}\cos\left(\Phi_{12}-\Phi_{32}\right)e^{-\alpha Z}.\label{eq:4_set_2} \,\,\,\,\,\,\,\,\,\,\,\,\,\,\,\,
\end{eqnarray}
For obtaining fixed points of the system given by (\ref{eq:3_set_1})-(\ref{eq:3_set_4}), we have to consider two separate cases: (1) CASE (i):  (\ref{eq:4_set_1})-(\ref{eq:4_set_2}) and  Condition I, and (2) CASE (ii):  (\ref{eq:4_set_1})-(\ref{eq:4_set_2}) and Condition II.

\subsection{CASE (i):}

Condition I can be further analyzed to produce 
\[
R_{12}=\frac{R_{32}}{\sqrt[]{R_{32}^{2}-1}}\,\,\textrm{and} \,\,R_{32}=\frac{R_{12}}{\sqrt[]{R_{12}^{2}-1}}. 
\]
Since $R_{12}$ and $R_{32}$ are real, this implies $R_{12} \in (1,\infty)$ as well as $R_{32} \in (1,\infty)$.

Subtracting   (\ref{eq:4_set_1}) from (\ref{eq:4_set_2}) we get
\begin{equation}
\frac{1}{R_{12}}\sin\left(\Phi_{32}\right)-\frac{1}{R_{32}}\sin\left(\Phi_{12}\right)=e^{-\alpha Z}\cos\left(\Phi_{12}-\Phi_{32}\right)\left[\left(\frac{1}{R_{12}}\right)^{2}-\left(\frac{1}{R_{32}}\right)^{2}\right].
\label{eq:5_1}
\end{equation}
Imposing Condition I   and $\dot{R}_{12}=0$ in (\ref{eq:3_set_1}), we obtain:
\begin{equation}
\frac{1}{R_{12}}\cos\left(\Phi_{32}\right)-\frac{1}{R_{32}}\cos\left(\Phi_{12}\right)=e^{-\alpha Z}\sin\left(\Phi_{12}-\Phi_{32}\right).
\label{eq:5_2}
\end{equation}
Note that imposing Condition I  and $\dot{R}_{32}=0$ in (\ref{eq:3_set_2}) also produces (\ref{eq:5_2}). Squaring and adding 
(\ref{eq:5_1}) and (\ref{eq:5_2}) and using Condition I, we obtain either

\subsubsection{CASE (i\,1):}

\begin{equation}
\frac{2}{R_{12}R_{32}} \cos\left(\Phi_{12}-\Phi_{32}\right)=1,
\end{equation}
\,\,\,\,\,\,\,\,\,\,\,or,
\subsubsection{CASE (i\,2):}
\begin{equation}
\frac{2}{R_{12}R_{32}} \cos\left(\Phi_{12}-\Phi_{32}\right)=e^{2\alpha Z}-1.
\end{equation}

For CASE\,(i\,1), using Condition I produces 
\begin{equation}
R_{12}=R_{32}=\sqrt{2} \,\,\mathrm{and}\,\, \Phi_{12}=\Phi_{32}=\sin^{-1}\left[\frac{1}{2\sqrt{2}}\left\{ e^{-\alpha Z}-\left(\gamma-2\right)e^{\alpha Z}\right\}\right].
\label{caseA1_soln}
\end{equation}

For CASE\,(i\,2), using Condition I produces (after a long but straight-forward algebra)
\begin{eqnarray}
& & \,\,\mathrm{Either}\,\, R_{12}=R_{32}=\sqrt{2} \,\,\mathrm{and}\,\, \Phi_{12}=\Phi_{32}=\sin^{-1}\left[\frac{1}{2\sqrt{2}}\left\{ e^{-\alpha Z}-\left(\gamma-2\right)e^{\alpha Z}\right\}\right]\nonumber \\
& & \,\,\mathrm{Or}\,\,  R_{12}=R_{32}=\sqrt{2} \,\,\mathrm{and}\,\,\Phi_{32}=\pi-\Phi_{12}=\sin^{-1}\left[\frac{1}{\sqrt{2}}\left\{ e^{\alpha Z}\pm\sqrt{1+\left(\gamma-1\right)e^{2\alpha Z}}\right\}\right].\,\,\,\,\,\,\,\,\,\,\,\,\,\,\,\,\,\,
\label{caseA2_soln}
\end{eqnarray}

The CASES (i\,1) and (i\,2) produce (\ref{caseA1_soln})-(\ref{caseA2_soln}) provided
\[
e^{-2\alpha Z}+2-2\sqrt{2}e^{-\alpha Z} \leq \gamma\leq e^{-2\alpha Z}+2+2\sqrt{2}e^{-\alpha Z}.
\] 

\subsection{CASE (ii):}
                                                                                     
Imposing Condition II   and $\dot{R}_{12}=0$ in (\ref{eq:3_set_1}), we obtain:
\begin{equation}
\cos\left(\Phi_{12}\right)\left[R_{32}e^{\alpha Z}-\frac{R_{32}}{R_{12}}\sin\left(\Phi_{12}\right)-\sin\left(\Phi_{32}\right)\right]=0
\label{eq:6_1}
\end{equation}
Note that imposing Condition II  and $\dot{R}_{32}=0$ in (\ref{eq:3_set_2}) also produces (\ref{eq:6_1}). From (\ref{eq:6_1}) and Condition II we get either

\subsubsection{CASE (ii\,1):}

\begin{equation}
 \cos\left(\Phi_{12}\right)=0\, \mathrm{and} \,\cos\left(\Phi_{32}\right)=0, \, \mathrm{hence} \,  \Phi_{12}=\pm \frac{\pi}{2}   \, \mathrm{and} \,   \Phi_{32}=\pm \frac{\pi}{2},
\end{equation}
\,\,\,\,\,\,\,\,\,\,\,or,
\subsubsection{CASE (ii\,2):}
\begin{equation}
\sin\left(\Phi_{32}\right)=R_{32}e^{\alpha Z}-\frac{R_{32}}{R_{12}}\sin\left(\Phi_{12}\right).
\end{equation}

CASE (ii\,1) can be divided into $4$ sub-cases:
\subsubsection{CASE\, (ii\,1.1):\,\,\, $\Phi_{12}=\frac{\pi}{2},\,\,\,\Phi_{32}=\frac{\pi}{2}$}

Subtracting (\ref{eq:4_set_1}) from (\ref{eq:4_set_2}) we obtain

\begin{equation}
\mathrm{(I)}\,\,R_{12}=\frac{R_{32}e^{-\alpha Z}}{R_{32}-e^{-\alpha Z}},\,\,\mathrm{or}\,\,\mathrm{(II)}\,\,R_{12}=R_{32}.
\end{equation}
When (I) holds, we find $\gamma=1-e^{-2\alpha Z}$. When (II) holds, $R_{12}$ and $R_{32}$ can be directly expressed in terms of 
$\gamma$ and $e^{\alpha Z}$:
\begin{equation}
R_{12}=R_{32}=\frac{1}{2}\left[e^{\alpha Z}\left(e^{-2\alpha Z}+2-\gamma\right)\pm\sqrt{e^{2\alpha Z}\left(e^{-2\alpha Z}+2-\gamma\right)^{2}-8}\right],
\label{eq:ss}
\end{equation}
provided \[
\gamma\leq e^{-2\alpha Z}+2-2\sqrt{2}e^{-\alpha Z}.
\] 
This basically implies $R_{12}$ and $R_{32}$ must be real, i.e.\ discriminant of (\ref{eq:ss}) is non-negative. 
\subsubsection{CASE\, (ii\,1.2):\,\,\, $\Phi_{12}=\frac{\pi}{2},\,\,\,\Phi_{32}=-\frac{\pi}{2}$}

Subtracting (\ref{eq:4_set_1}) from (\ref{eq:4_set_2}) we obtain
\begin{equation}
R_{12}=\frac{R_{32}e^{-\alpha Z}}{R_{32}+e^{-\alpha Z}}\,\,\,\mathrm{and}\,\,\,\gamma=1-e^{-2\alpha Z}.
\end{equation}
\subsubsection{CASE\, (ii\,1.3):\,\,\, $\Phi_{12}=-\frac{\pi}{2},\,\,\,\Phi_{32}=\frac{\pi}{2}$}
Subtracting (\ref{eq:4_set_1}) from (\ref{eq:4_set_2}) we obtain
\begin{equation}
R_{32}=\frac{R_{12}e^{-\alpha Z}}{R_{12}+e^{-\alpha Z}}\,\,\,\mathrm{and}\,\,\,\gamma=1-e^{-2\alpha Z}.
\end{equation}
\subsubsection{CASE\, (ii\,1.4):\,\,\, $\Phi_{12}=-\frac{\pi}{2},\,\,\,\Phi_{32}=-\frac{\pi}{2}$}
\begin{equation}
R_{12}=R_{32}=\frac{1}{2}\left[-e^{\alpha Z}\left(e^{-2\alpha Z}+2-\gamma\right)\pm\sqrt{e^{2\alpha Z}\left(\gamma-2-e^{-2\alpha Z}\right)^{2}-8}\right],
\label{eq:ss2}
\end{equation}
provided \[
\gamma\geq e^{-2\alpha Z}+2+2\sqrt{2}e^{-\alpha Z}.
\] 
Like CASE (ii 1.1), CASE (ii 1.4) is also valid when $R_{12}$ and $R_{32}$ are real, i.e.\ discriminant of (\ref{eq:ss2}) is non-negative. 

\setcounter{subsubsection}{0}

\subsubsection{CASE\, (ii\,2):\,\,\, $\sin\left(\Phi_{32}\right)=R_{32}e^{\alpha Z}-\frac{R_{32}}{R_{12}}\sin\left(\Phi_{12}\right)$}

This condition, along with Condition II when substituted in (\ref{eq:4_set_1}) produces $\gamma=1-e^{-2\alpha Z}$.

In summary, from Conditions I and II and CASES (i) and (ii),  we obtain  (\ref{eq:casei})-(\ref{eq:caseiii}), provided we ignore the singular case when  $\gamma=1-e^{-2\alpha Z}$. This particular case is interesting in its own right and will be addressed in a future communication.

  \bibliographystyle{jfm}
  \bibliography{paper1}
  




 


\end{document}